\documentclass[preprint,showpacs,preprintnumbers,amsmath,amssymb]{revtex4}

\usepackage{graphicx}
\usepackage{dcolumn}
\usepackage{bm}

\begin{document}

\preprint{}

\title{Pressure-induced topological phase transition in polar-semiconductor BiTeBr}

\author{Ayako Ohmura$^1$}%
  \email{ohmura@phys.sc.niigata-u.ac.jp}%
\author{Yuichiro Higuchi$^2$, Takayuki Ochiai$^3$, Manabu Kanou$^4$, \\ Fumihiro Ishikawa$^{1,3}$, Satoshi Nakano$^5$, Atsuko Nakayama$^{1,6}$, \\ Yuh Yamada$^{1,3}$, Takao Sasagawa$^4$}%

\affiliation{$^{1}$ Center for Transdisciplinary Research, Niigata University, 8050 Ikarashi 2-no-cho, Nishi-ku, Niigata, Niigata 950-2181, Japan}%

\affiliation{$^{2}$ Graduate School of Science and Technology, Niigata University, 8050, Ikarashi 2-no-cho, Nishi-ku, Niigata, 950-2181, Japan}%

\affiliation{$^{3}$ Faculty of Science, Niigata University, 8050, Ikarashi 2-no-cho, Nishi-ku, Niigata, 950-2181, Japan}%

\affiliation{$^{4}$ Laboratory for Materials and Structures, Institute of Innovation Research, Tokyo Institute of Technology, 4259 Nagatsuta, Midori-ku, Yokohama, Kanagawa, 226-8503, Japan}%

\affiliation{$^{5}$ Ultra-High Pressure Processes Group, National Institute for Materials Science (NIMS), 1-1 Namiki, Tsukuba, Ibaraki, 305-0044, Japan}%

\affiliation{$^{6}$ Department of Physical Science and Materials Engineering, Faculty of Science and Engineering, Iwate University, 4-3-5 Ueda, Morioka 020-8550, Japan}%

\date{\today}

\begin{abstract}
We performed X-ray diffraction and electrical resistivity measurement up to pressures of 5 GPa and the first-principles calculations utilizing experimental structural parameters to investigate the pressure-induced topological phase transition in BiTeBr having a noncentrosymmetric layered structure (space group $P3m1$). The $P3m1$ structure remains stable up to pressures of 5 GPa; the ratio of lattice constants, $c/a$, has a minimum at pressures of 2.5 - 3 GPa. In the same range, the temperature dependence of resistivity changes from metallic to semiconducting at 3 GPa and has a plateau region between 50 and 150 K in the semiconducting state. Meanwhile, the pressure variation of band structure shows that the bulk band-gap energy closes at 2.9 GPa and re-opens at higher pressures. Furthermore, according to the Wilson loop analysis, the topological nature of electronic states in noncentrosymmetric BiTeBr at 0 and 5 GPa are explicitly revealed to be trivial and non-trivial, respectively. These results strongly suggest that pressure-induced topological phase transition in BiTeBr occurs at the pressures of 2.9 GPa.
\end{abstract}

\maketitle

\section{\label{sec:level1}Introduction}
The spin-orbit interaction (SOI) induces a variety of interesting phenomena. A topological insulator, which is a novel electronic state having metallic surface and insulating bulk states, is a representative example and is found in many bismuth compounds due to the strong atomic SOI of bismuth. The Rashba-type spin-split is another SOI-induced phenomena. Polar semiconductors BiTe$X$ ($X$ = Cl, Br, I) are well known as the materials exhibiting the giant bulk-Rashba spin-split, which is caused by the combination of a noncentrosymmetric crystal structure and strong SOI \cite{Bahramy2011, Ishizaka2011, YLChen2013, Kanou2013}. Among BiTe$X$, only BiTeCl is reported to be an inversion asymmetric topological insulator at ambient conditions \cite{YLChen2013}.

Recently, a pressure-induced topological phase transition in BiTeI was theoretically predicted by Bahramy~\textit{et al.} \cite{Bahramy2012}. In BiTeI, the Rashba-type spin-split occurs along the $L$-$A$-$H$ direction. The conduction and valence bands near Fermi energy are dominantly composed of Bi-6$p$ and Te, I-5$p$ orbitals, respectively. They reported that the electronic states of BiTeI change from a trivial semiconductor to a topological insulator by band inversion of $p_z$-orbitals, which are originated in Bi, Te, and I, above a critical pressure of $P_c$ {$\sim$} 1.7 - 4.1 GPa under hydrostatic condition. Since the publication of Bahramy \textit{et al.} \cite{Bahramy2012}, high-pressure experimental and theoretical studies have been performed in BiTeI \cite{Posonov2013, Tran2014, YChen2013, Xi2013, Park2015}. From these studies, there has been a suggestion for the occurrence of a topological phase transition. According to some studies \cite{Posonov2013, YChen2013, Xi2013}, the pressures of $P$ {$\sim$} 2 - 4.5 GPa, at which the ratio of lattice constants has an extremum, are consistent with those at which the band-gap closing and a maximum in free-carrier spectral weight are observed. It suggests that a change in the electronic state correlates with the ratio of lattice constants. 

Until recently, the crystal structure of BiTeBr had been thought to be disordered Te/Br sites with $P\bar{3}m1$ symmetry (No. 164) \cite{Shevelokv1995}. However, recent studies show that the Rashba-type spin-split also exists in the band structure of BiTeBr \cite{Sakano2013, Akrap2014, Eremeev2013}, and it is now considered proof of having the same ordered structure as BiTeI with space group $P3m1$ (No.156) \cite{Sakano2013}. Its crystal structure at ambient conditions is a noncentrosymmetric layered structure along the $c$-axis as shown in an inset of Fig.~\ref{fig:resistivity}. A unit of Te-Bi-Br three layers is weakly bonded by van der Waals forces along the $c$-axis. In contrast, Bi-Te and Bi-Br bonds in this unit have covalent and ionic properties, respectively. Due to the structural and electronic similarities between the two materials, we could expect the topological phase transition to occur in BiTeBr under high pressure. Furthermore, scanning tunneling microscopy revealed the distribution of submicron-scale domains composed of $p$- and $n$-type semiconducting domains with opposite stacking sequences in BiTeI \cite{Kohsaka2015}. Since BiTeBr has no such domains \cite{Fiedler2015}, it is not necessary to consider its effect on evaluating transport properties under high pressure. 
 
The purpose of this study is to investigate the pressure-induced topological phase transition in BiTeBr experimentally. Though there are various high-pressure studies of BiTeI as mentioned above, a variation of transport properties associated with the phase transition have not been reported yet. We, therefore, performed the high-pressure X-ray diffraction and electrical resistivity measurements, and the first-principles calculations using experimental structural parameters. Furthermore, since the crystal structure of BiTeBr is noncentrosymmetric, it is not possible to calculate its $Z_2$ topological invariants directly from the parity analysis as described in Ref.~\cite{Fu2007}. We, therefore, performed the Wilson loop analysis to evaluate topologically distinct properties of electronic states.

\section{Experimental}
The powder sample for X-ray diffraction was prepared as follows. The starting material was a mixture of high-purity elemental Bi (5N), Te (5N), and BiBr$_3$ (4N) at a molar ratio of 2:3:1, and put in a quartz tube. The sealed quartz tube was heated up to 800 {$^\circ$}C and cooled to 400 {$^\circ$}C over 100 hours, resulting in an ingot of BiTeBr precursor. In a nitrogen-filled glove box, this ingot was ground into powder for 1 hour using an agate mortar. The powder was sealed in a quartz tube under nitrogen atmosphere and then heated at 300 {$^\circ$}C over 100 hours. Single crystals of BiTeBr were grown utilizing a modified self-flux technique \cite{Kanou2013}. The starting material was a mixture of high-purity elemental Bi (5N), Te (5N), and BiBr$_3$ (4N) at a molar ratio of 2:3:4, in which the excess BiBr$_3$ serves as the self-flux. A quartz tube with the starting material was evacuated and sealed. Crystal growth was carried out using a horizontal two-zone Bridgman furnace. First, both zones were heated up to 800 {$^\circ$}C and kept stable for 10 hours to react the starting material completely. Then, one zone was decreased to 750 {$^\circ$}C. Keeping the temperature difference of 50 {$^\circ$}C between the two zones, over 100 hours, they were slowly cooled to 300 and 250 {$^\circ$}C, respectively.

Electrical resistivity measurements under high pressures and low temperatures were performed using the Cu-Be modified Bridgman anvil cell \cite{Nakanishi2002, Ishikawa2007}. Samples were cut out of single crystals of BiTeBr; the sizes of samples were 0.30${\times}$0.75${\times}$0.070 (sample 1) and 0.084${\times}$0.31${\times}$0.14 (sample 2) mm$^3$, respectively. The sample was placed into a Teflon capsule together with a mixture of Fluorinert (FC70:FC77 = 1:1), which is a pressure-transmitting medium used for quasi-hydrostatic compression. The high-pressure cell was compressed at room temperature and then cooled down to {$\sim$} 3 K by Gifford-McMahon cryogenic refrigerator (Iwatani Industrial Gases Corporation). The value of pressure generated in the capsule against a load on the high-pressure cell was calibrated in advance by the critical pressures of structural phase transitions in elemental bismuth (5N) as a standard material. Temperature dependence of resistivity was measured under high pressures up to 5 GPa. 

For X-ray diffraction, we used a diamond anvil cell (DAC) with a culet of 0.45 mm in diameter. The powder sample was placed in a sample chamber, 180 $\mathrm{{\mu}}$m in diameter and 60 $\mathrm{{\mu}m}$ thick, made by drilling a small hole in a SUS301 gasket. All handling of powder sample was performed in an argon atmosphere to avoid the reaction with moisture in the air. For hydrostatic compression, the sample hole was filled with helium fluid (He), which was compressed to {$\sim$} 180 MPa, as a pressure-transmitting medium \cite{Takemura2001}. The DAC was sealed at a pressure of 0.58 GPa. X-ray diffraction with synchrotron radiation was carried out at beamline AR-NE1A of Photon Factory in High-Energy Accelerator Research Organization (KEK) in Tsukuba, Japan. An incident beam was tuned to the energy of 29.77 keV ($\lambda$ = 0.4165 \AA) and a size of 30 $\mathrm{{\mu}m}$ in diameter. An imaging plate was used as a detector (Rigaku R-AXIS IV). We measured diffraction patterns under high pressures up to 4.93 GPa at room temperature. The experimental values of pressure were determined using the ruby fluorescence method \cite{Zha2000}. The Rietveld analysis was performed using RIETAN-FP to obtain structural parameters \cite{Rietveld1969, Izumi2007}.

First-principles calculations in the framework of density functional theory (DFT) were performed within the generalized gradient approximation (GGA), using the pseudopotential plane-wave method as implemented in the ABINIT code \cite{Gonze2009}. To perform relativistic calculations, norm-conserving Hartwigsen-Goedecker-Hutte (HGH) pseudopotentials were employed \cite{Hatwigsen1998}. Electronic structures were calculated using structural parameters, which were experimentally obtained at each pressure in this study. Within the experimental unit cell, atomic coordinates were optimized based on the minimum of force on each atom. The topological $Z_2$ invariants were evaluated using the Wilson loop method as described in Ref. \cite{Soluyanov2011}. 

\section{Results}

\begin{figure*}
\includegraphics[height=8 cm]{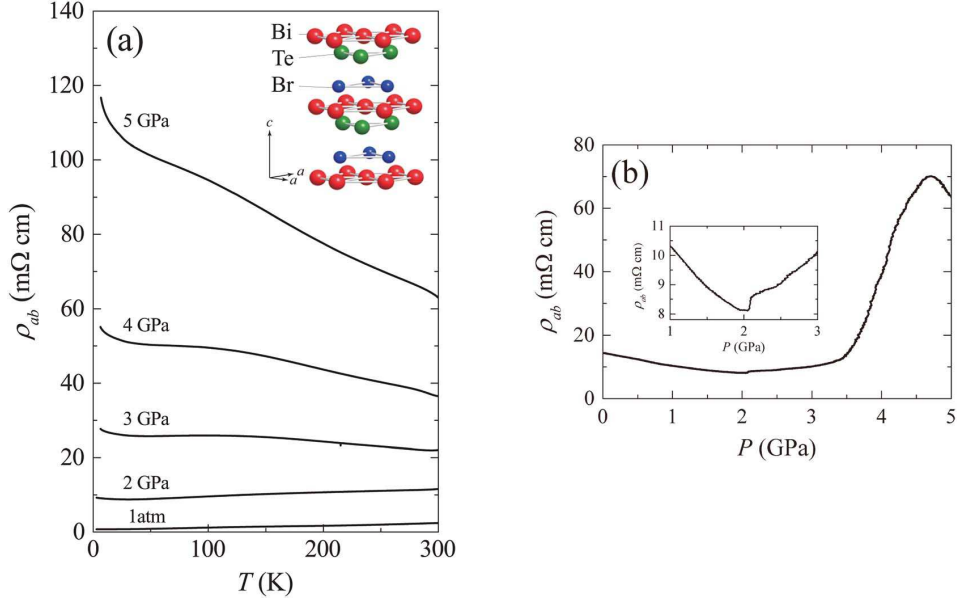}
\caption{\label{fig:resistivity} (Color online) (a) Temperature dependence of electrical resistivity at pressures up to 5 GPa in a BiTeBr single crystal. Inset: crystal structure of BiTeBr at ambient conditions. (b) Variation of electrical resistivity as a function of pressure at ambient temperature. Inset: magnified view of pressure ranges of 1 to 3 GPa.}
\end{figure*}

Figure~\ref{fig:resistivity}(a) shows the temperature dependences of electrical resistivity in BiTeBr (sample 1) with increasing pressure. The resistivity has the value of {$\sim$} 2.4 m$\Omega$ cm at ambient conditions and decreases with decreasing temperature similar to a previous paper \cite{Kulbachinskii}. The metallic-like behavior in BiTe$X$ is thought to be the result of self-doping caused by the antisite or vacancy defects, though the band structure of BiTeBr demonstrates a semiconductor in the case of a perfect crystal \cite{Kulbachinskii}. By applying pressure, the temperature coefficient of resistivity changes from positive to negative between 2 and 3 GPa. Furthermore, we can see a plateau region between 50 and 150 K in the semiconducting phase above 3 GPa. The plateau region is suppressed with increasing pressure, while the semiconducting behavior is enhanced.

The experimental values of pressure in sample 1 have a margin of error of {$\pm$}1 GPa, because the load applied to the cell was maintained utilizing the clamp screw of the cell before the cooling cycle. In the pressure calibration procedure, on the other hand, the standard material (elemental bismuth) is continuously pressurized. To precisely apply the calibrated pressure values and make clearer the relation to the resistivity behavior, we measured sample 2 through the same process as the pressure calibration. Figure~\ref{fig:resistivity}(b) shows the result of sample 2, which was pressurized continuously at a constant rate at room temperature. The resistivity starts decreasing and reaches a minimum at 2.1 GPa. Beyond this pressure, a slight jump is observed as shown in the inset. On further compression, the resistivity increases, showing a particularly rapid increase above 3.5 GPa, and then reaches a maximum value at 4.7 GPa.

\begin{figure*}
\includegraphics[height=9 cm]{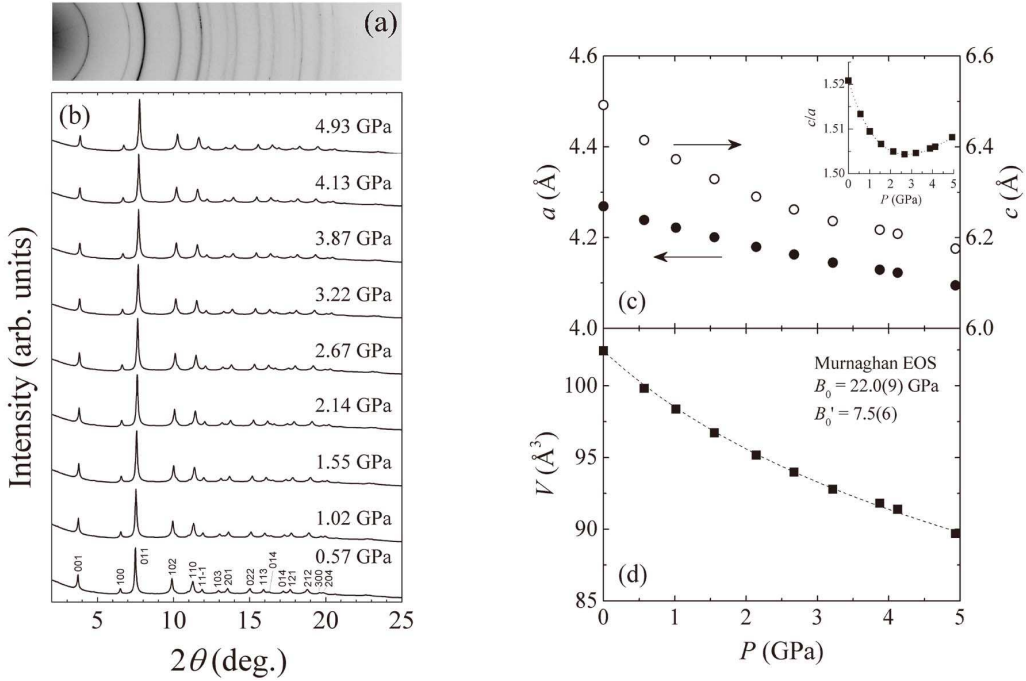}
\caption{\label{fig:XRD} (a) X-ray diffraction image of the polycrystalline BiTeBr sample at 0.57 GPa. (b) Diffraction patterns on compression at ambient temperature. (c) Pressure variations of lattice constants (main panel) and their ratio $c/a$ (inset). (d) Compression curve of BiTeBr. Dashed line represents the Murnaghan EOS fit to the experimental data.}
\end{figure*}

Figure~\ref{fig:XRD}(a) shows an exemplary X-ray diffraction image obtained at 0.57 GPa. The image has several spots along Debye-Scherrer rings, indicating the existence of single crystalline grains of a certain size. This is because the polycrystalline sample was not ground to avoid decomposition after final annealing in the synthesis process. Diffraction patterns measured up to a pressure of 4.93 GPa are shown in Fig.~\ref{fig:XRD}(b). The $P3m1$ symmetry indexes all reflections obtained at 0.57 GPa; lattice constants are $a$ = 4.2388(5) \AA~and $c$ = 6.4145(10) \AA. Pattern profiles show no change with an increase in pressure, indicating that the $P3m1$ structure is stable at all pressures below 4.93 GPa. 

Figures~\ref{fig:XRD}(c) and (d) show pressure variations of lattice constants and volumes, $V$, in the $P3m1$ structure, respectively. Lattice constants and volume monotonically decrease with pressure. On the other hand, the ratio of lattice constants, $c/a$, reaches a minimum at pressures of 2.5 - 3 GPa as shown in the inset of Fig.~\ref{fig:XRD}(c). These behaviors and the values of $a$ and $c$ clearly reproduce those of our previous work on BiTeBr \cite{Ohmura}. The compression curve of $V$ is fitted by Murnaghan's equation of state \cite{Murnaghan1944}:
\begin{equation}
P=B_0/B^{'}_0[(V_0/V)^{B^{'}_0}-1],
\end{equation}
where $V_0$ and $V$ are volumes at ambient and high pressures, $P$ is in units of GPa, $B_0$ is the bulk modulus, and $B^{'}_0$ is the pressure derivative. The fit gives parameters of $B_0$ = 22.0(9) GPa and $B_0^{'}$ = 7.5(6), respectively, for BiTeBr with the $P3m1$ structure.

Figures~\ref{fig:band gap}(a) and (b) show the pressure variations of the bulk electronic states near Fermi energy, $E_F$, along the $L$-$A$-$H$ direction. At ambient pressure (red curves), the Rashba-type spin-split and the band gap are clearly observed. The Rashba parameter is {$\alpha_\mathrm{R}$ = $2E_\mathrm{R}$/$k_\mathrm{R}$, where $k_\mathrm{R}$ and $E_\mathrm{R}$ are the momentum offset of conduction band minimum and the Rashba energy, respectively. The {$\alpha_\mathrm{R}$} of bulk conduction bands at around the $A$ point are 3.5 eV \AA~  in the $A$-$L$ direction and 4.6 eV \AA~ in the $A$-$H$ direction. The band-gap energy, $E_g$, at ambient pressure is about 0.17 eV. $E_g$ monotonously decreases up to a pressure of 2.90 GPa (green curves) and then turns into a continuous increase beyond this point. Energy-band dispersions near $E_F$ transform from parabolic to linear upon approaching zero energy-gap. 

Figure 3(c) shows the pressure variation of the values of $E_g$ (red circles) obtained at each pressure. To estimate a critical pressure, $P_c$, at which the band-gap closes, we inverted the values of $E_g$ against the zero-gap line, creating a mirror image (blue circles). The pressure variation of red circles smoothly connects to that of blue circles. We can, therefore, make an estimate of the critical pressure $P_c$ = 2.9 GPa for the band-gap closing since both variations intersect on the line of $E_g$ = 0 eV at this pressure. The value of $P_c$ is consistent with the pressure range at which the change in the temperature coefficient of resistivity and the minimum value of $c/a$ are observed.

Recently, the high-pressure work in BiTeBr was reported in detail \cite{Sans2016}. In this report, the magnitude of band-gap obtained by the DFT calculations decreases as a function of pressure, and its closing, however, is not observed. We, on the other hand, obtained the reproducibility of the band-gap closing at ~3 GPa by calculations using two sets of experimental data in this study and Ref. \cite{Ohmura}. With regards to the difference between two works, we have no clear answer yet. One of the possibilities seems to be the difference of lattice constant $c$ and atomic positions after the optimization of structural parameters. In this study, the atomic positions were optimized under fixed experimental lattice constants. The optimized atomic parameters are 1$b$ (1/3, 2/3. 0.29878) for Br and 1$c$ (2/3, 1/3, 0.72648) for Te at ambient conditions, which are {$\sim$} 10 and {$\sim$} 3 {\%} higher than those values in Ref. \cite{Sans2016}, respectively.

\begin{figure}
\includegraphics[height=10cm]{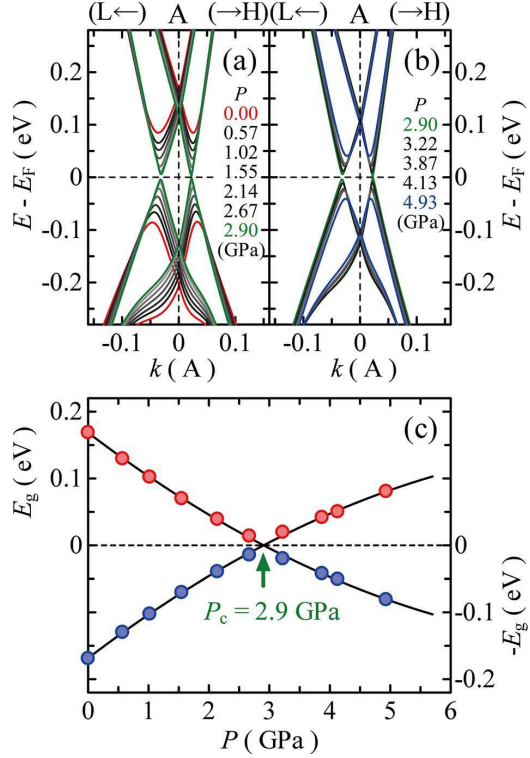}
\caption{\label{fig:band gap}(Color online) (a) Bulk electronic states of BiTeBr along the $L$-$A$-$H$ direction calculated using experimental lattice constants under various pressures. (b) Pressure variation of energy gap ($E_g$) shown as red circles. Blue circles are symmetric points of red ones with respect to $E_g$ = 0 eV. Solid lines are guide for the eye.}
\end{figure}

As mentioned above, since the crystal structure of BiTeBr is noncentrosymmetric, it is not possible to calculate its $Z_2$ topological invariants directly from the parity analysis as described in Ref.~\cite{Fu2007}. We, therefore, performed the Wilson loop analysis to evaluate topologically distinct properties of electronic states above and below $P_c$. In this analysis, the equivalent information to the $Z_2$ topological invariant is obtained from the evolution of the Wannier charge center (WCC), which is calculated around a closed loop in the Brillouin zone and depends on the bulk wave function \cite{Soluyanov2011, Fu2006, Taherinejad2014}. Figure~\ref{fig:WCC} shows the results of its analysis in BiTeBr at pressures of 0 and 5 GPa; a pair of the WCC evolution around $y_n$ = 0 is highlighted as blue and red curves. The degeneration and separation of the WCC pair at ($k_x/\pi$, $k_z/\pi$) = (0, 1) under 0 and 5 GPa, respectively, indicate topologically trivial and non-trivial electronic states. The WCC evolution reflects the surface energy band, and the calculated result indicates the band inversion near the $A$ point. 

\begin{figure}
\includegraphics[height=10cm]{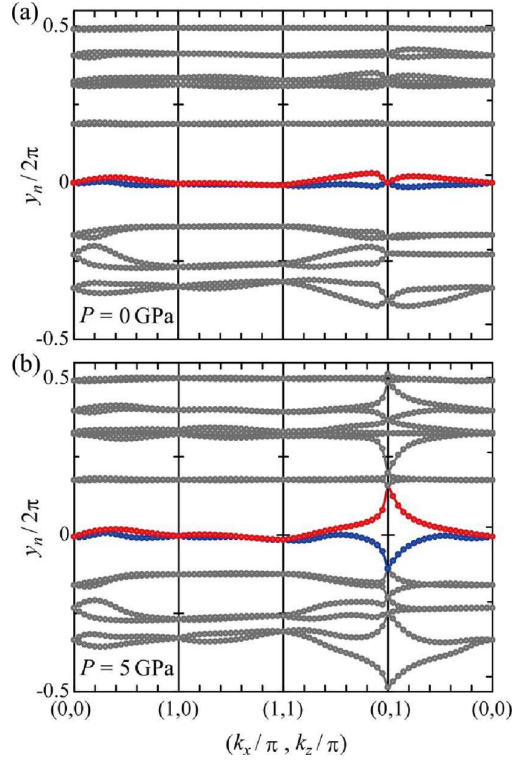}
\caption{\label{fig:WCC} (Color online) Wilson loop analysis of topological electronic states in BiTeBr at pressures of (a) 0 GPa and (b) 5 GPa. A pair of the Wannier charge center evolution around $y_n$ = 0 is highlighted as blue and red curves.}
\end{figure}

\section{Discussion}

The results of calculations shown in Figs.~\ref{fig:band gap} and~\ref{fig:WCC} strongly suggest that the topological phase transition occurs at the $P_c$ of 2.9 GPa in BiTeBr. First, we will explain the change in the temperature dependence of resistivity. The transport property becomes semiconducting beyond 3 GPa, and its behavior develops with an increase in pressure. As compared to the pressure variation of $E_g$, it is suggested that the localization of bulk carriers progresses with an increase in the magnitude of $E_g$. The existence of the plateau region also supports the formation of the topologically non-trivial phase having the metallic surface state above $P_c$. A similar behavior is observed in the topological insulators \cite{Taskin2000}, consistent with a picture that thermally excited carriers in the bulk are suppressed upon decreasing temperature and the surface conduction becomes dominant at low temperatures.

Furthermore, the pressure variation of $E_g$ is reflected in that of electrical resistivity at room temperature as shown in Fig. 1(b). The value of resistivity shows a discontinuous increase at 2.1 GPa. Concerning this change, we infer that the minimum value and subsequent jump are caused by the closing and re-opening of $E_g$, respectively. The experimental pressure of anomaly in resistivity is lower than the computed critical pressure of $E_g$, $P_c$ = 2.9 GPa. We think that this difference depends on the degree of hydrostaticity in each experiment and the magnitude of experimental errors in resistivity measurement. The phase transition tends to occur at lower pressures as the degree of hydrostaticity degrades, in general. Since the hydrostaticity in the Bridgman anvil cell with liquid medium (resistivity) is lower than that in the diamond anvil cell with He medium (X-ray diffraction), the resistivity minimum was observed at lower pressures than the band-gap closing estimated from X-ray diffraction.

Since the $P3m1$ structure stably exists up to the pressure of 5 GPa, our results indicate that the bulk electronic state changes from metallic to semiconducting under the same structure. There are a few reports on the transitions from the metallic phase to the semiconducting/insulating phase unaccompanied by the structural change under high pressure. One example is one-dimensional platinum complexes \cite{Takeda2000, Shirotani1991}, which are trivial insulators at ambient pressure. They exhibit pressure-induced insulator-to-metal-to-insulator transitions with the resistivity minimum at room temperature. Although the detail of transition mechanism of platinum complexes has not been elucidated yet, the displacement of top two valence bands and the separation from a conduction band with pressure were discussed \cite{Takeda2000}. Unlike BiTeBr, the pressure variations of resistivity in platinum complex, however, show a smooth change without a discontinuous jump. Therefore, we think that the small jump of resistivity observed in this study is unique to BiTeBr, namely its topological phase transition.

The band-gap closing in BiTeBr accompanies the minimum of $c/a$. In layered compounds in Bi-system such as Bi$_2$Te$_3$ and Bi$_2$Se$_3$, the minimum of $c/a$ at high pressures is often discussed in relation to the Lifshitz transition, in which the Fermiology (i.e., the topology of the Fermi surface) changes. The Lifshitz transition is evaluated utilizing the Eulerian strain $f_E$ and the reduced pressure $H$ (= $B_0$ + (3/2) $B_0$ ($B_0'$-4) $f_E$ ) \cite{Jacobsen2007, Vilaplana2011, Polian2011}; the slope of $H$ as a function of $f_E$ changes when the Lifshitz transition occurs. In BiTeBr, there is no such noticeable change at $P_c$ = 2.9 GPa, at which the band-gap closes. Therefore, it indicates that the electronic transition we observed is not the Lifshitz transition.

According to Bahramy \textit{et al.} \cite{Bahramy2012}, the band inversion between Te, I-5$p_z$ and Bi-6$p_z$ orbitals in BiTeI is one of the key factors in its topological phase transition. In BiTeI, when the $c/a$ of $P3m1$ structure reaches a minimum at $P$ {$\sim$} 2.0-2.9 GPa, a maximum in free-carrier spectral weight is observed \cite{Xi2013}. Additionally, a recent experiment reported that the unusual size increase of inner Fermi surface (FS) in Rashba bands and the curvature change of outer FS are observed above 2 GPa \cite{Park2015}, suggesting the correlation between the electronic state and the $c/a$-minimum. In consideration of these results in BiTeI, we infer that the topological phase transition in BiTeBr also occurs at the minimum of $c/a$, in other words, the maximum distortion of crystalline lattice along the $c$-axis causes the band inversion of $p_z$-orbitals.

\section{Summary}

We investigated pressure-induced topological phase transition in BiTeBr by combining experimental and theoretical studies. The transport property changes from metallic to semiconducting behavior between 2 and 3 GPa. The $P3m1$ structure remains stable up to pressures of 5 GPa, and its $c/a$ has a minimum at pressures of 2.5 - 3 GPa. The $E_g$, which is calculated using the experimental structural parameters, closes and successively re-opens at the pressure of $P_c$ = 2.9 GPa, suggesting the occurrence of topological phase transition. The semiconducting behavior above 3 GPa is considered to be due to the localization of bulk carriers resulting from this re-opening of $E_g$. Furthermore, the Wilson loop analysis clearly shows the different topological states above and below $P_c$: a trivial and a non-trivial states at 0 and 5 GPa, respectively. We, therefore, concluded that the topological phase transition in BiTeBr occurs at the pressure of 2.9 GPa, accompanied by the $c/a$-minimum in the $P3m1$ structure.

\begin{acknowledgments}
The present work was performed under Proposal No. 2012G162 and 2014G107 of Photon Factory, KEK. This work was supported by Grant-in-Aid for Young Scientists (B) (No. 24740229) and for Scientific Research (B) (Nos. 16H03847 and 24340078), and the "Topological Quantum Phenomena" (No. 25103710) Grant-in-Aid for Scientific Research on Innovative Areas from MEXT of Japan, and Collaborative Research Projects 2013-2015 in Materials and Structures Laboratory, Tokyo Institute of Technology.
\end{acknowledgments}

\newpage

\end{document}